\documentstyle[aps,prl,multicol]{revtex}

\def\01{\{0,1\}}
\def\x{\times}
\def\Q{\mbox{\sf q}}
\def\ket#1{\mbox{$| #1 \rangle$}}
\def\ox{\otimes}
\def\o{\hspace*{-0.6mm}\cdot\hspace*{-0.3mm}}
\def\1n{\{1,\ldots,n\}}
\def\one{\mbox{\boldmath $1$}}
\def\nl{\newline}
\def\ni{\noindent}
\def\ii{\hspace*{8mm}}
\def\ee{\vspace*{3mm}}
\def\loud#1{\noindent{\bf #1 }}
\def\half{\textstyle{1 \over 2}}

\begin{document}

\title{Substituting Quantum Entanglement for Communication}

\author{Richard Cleve$^{1}$\thanks{\tt cleve@cpsc.ucalgary.ca} 
and 
\addtocounter{footnote}{3}
Harry Buhrman$^{2}$\thanks{\tt buhrman@cwi.nl}}
\address{$^1$ Department of Computer Science, University of Calgary, 
Calgary, Alberta, Canada T2N 1N4 \\
$^2$ CWI, P.O. Box 94070, 1090 GB Amsterdam, The Netherlands}

\draft

\maketitle

\begin{abstract}
We show that quantum entanglement can be used as a substitute for 
communication when the goal is to compute a function whose input data 
is distributed among remote parties.
Specifically, we show that, for a particular function among three parties 
(each of which possesses part of the function's input), a prior quantum 
entanglement enables one of them to learn the value of the function with 
only two bits of communication occurring among the parties, whereas, without 
quantum entanglement, three bits of communication are necessary.
This result contrasts the well-known fact that quantum entanglement cannot 
be used to simulate communication among remote parties.
\end{abstract}


\begin{multicols}{2}[]

If a set of entangled particles are individually measured, the resulting 
outcomes can exhibit ``nonlocal'' effects \cite{EPR,Bell,CHSH,GHZ,Mermin}.
These are effects that, from the perspective of ``classical'' physics, 
cannot occur unless ``instantaneous communications'' occur among 
the particles, which convey information about each particle's measurement 
to the other particles.

On the other hand, no communication actually occurs among the entangled 
particles when they are measured.
To phrase this in operational terms, entangled particles cannot be used to 
simulate communication.
For example, if two physically separated parties, Alice and Bob, 
initially possess particles whose quantum states are entangled and then 
Bob obtains a bit of information $x$, there is no operation 
that Bob can apply to his particles that will have the effect of conveying 
$x$ to Alice when she performs measurements on her particles.
Moreover, entanglement cannot even be used to {\em compress} information: 
for Bob to convey $n$ bits (with arbitrary values) to Alice, he must send $n$ 
bits---sending $n-1$ bits will not suffice.
Also, similar results apply to communications involving more than two 
parties.

Consider the following related but different scenario.
Alice obtains an $n$-bit string $x$, and Bob obtains an $n$-bit string 
$y$ and the goal is for Alice to determine $f(x,y)$, for some function 
$f : \01^n \x \01^n \rightarrow \01$, with as little communication between 
Alice and Bob as possible.
This can always be accomplished by Bob sending his $n$ bits to Alice, but 
fewer bits may suffice.
For example, for the function 
\begin{equation}
f(x,y) = x_1 + \cdots + x_n + y_1 + \cdots + y_n
\end{equation}
(where $+$ means addition modulo two), it suffices for Bob to send 
a single bit (namely, $y_1 + \cdots + y_n$) to Alice.
On the other hand, for other functions, such as the {\em inner product} 
(in modulo two arithmetic)
\begin{equation}
f(x,y) = x_1 \o y_1 + \cdots + x_n \o y_n,
\end{equation}
$n$ bits of communication turn out to be necessary (see \cite{KN} for 
a proof of this).
Thus, even though the goal is for Alice to acquire a single bit of 
information, this bit depends on the $2n$ bits distributed among Alice 
and Bob in such a way that they must exchange $n$ bits between 
them in order for Alice to determine this bit.
For a function $f : \01^n \x \01^n \rightarrow \01$, the minimum number 
of bits that must be communicated between Alice and Bob in order for 
Alice to determine $f(x,y)$ is called the {\em communication complexity} 
of $f$.
Several aspects of communication complexity are surveyed in \cite{KN}.

The question that we consider is whether or not a prior quantum 
entanglement can reduce communication complexity.
For example, if Alice and Bob initially possess entangled particles, 
can they compute some functions using less communication than would 
be required without the entangled particles?
Although we do not presently know the answer for this two-party scenario, 
we exhibit an analogous three-party scenario where entanglement {\em does} 
reduce communication complexity.
The function is based on Mermin's version \cite{Mermin} of 
``Bell-nonlocality without probabilities''.

Consider the following three-party scenario.
Alice, Bob, and Carol receive $n$-bit strings $x$, $y$, and $z$ 
respectively, which are subject to the condition that 
\begin{equation}
x + y + z = \one,
\label{conds}
\end{equation}
where $+$ is applied bitwise (modulo two) and 
$\one = \overbrace{11 \ldots 1}^n\,$.
The goal is for Alice to determine the value of 
\begin{equation}
f(x,y,z) = x_1 \o y_1 \o z_1 + \cdots + x_n \o y_n \o z_n.
\label{fcn}
\end{equation}
An alternative way of expressing this problem is to impose no restriction 
on the inputs, $x$, $y$, $z$, and to extend $f$ to a {\em relation} such 
that on the points where Eq.~(\ref{conds}) is violated, both 0 and 1 are 
acceptable outputs.
Clearly, this problem has the same communication complexity as the 
original one.

We show that, for the cases where $n \ge 3$:
\begin{itemize}
\item without a prior entanglement, three bits of communication are 
{\em necessary} for Alice to determine $f(x,y,z)$; and 
\item with a certain prior entanglement, two bits of communication are 
{\em sufficient} for Alice to determine $f(x,y,z)$.
\end{itemize}
Thus, even though entanglement cannot be used to simulate communication, 
it can nevertheless act as a {\em substitute} for communication when 
the goal is to compute a function with distributed data.
We also show that the lower bound of three in the case of no entanglement 
cannot be improved.
This is done by exhibiting a three-bit protocol.

Recently, Grover \cite{Grover} has independently demonstrated that 
quantum entanglement can reduce communication complexity in a different 
context.\ee

\loud{A TWO-BIT QUANTUM PROTOCOL:}\ee

We now show that if A(lice), B(ob), and C(arol) initially 
share a certain entanglement of qubits then there is a protocol in which B 
and C each send a single bit to A, which enables A to determine $f(x,y,z)$ 
(as defined by Eqs.~(\ref{conds}) and (\ref{fcn})).

The entanglement involves $3n$ qubits, with each party having $n$ of them.
Call the $n$ qubits that party $p \in \{A,B,C\}$ starts with 
$\Q^p_1, \ldots, \Q^p_n$.
For each $i \in \1n$, let the triple $\Q^A_i \Q^B_i \Q^C_i$ be in state 
\begin{equation}
\half (\ket{001} + \ket{010} + \ket{100} - \ket{111}).
\label{state}
\end{equation}
(This is equivalent to the state examined in \cite{Mermin} but in 
an alternate basis.)
For convenience, in this section, we write $x^A$, $x^B$, and $x^C$ for 
the inputs of A, B, and C, instead of $x$, $y$, and $z$, 
respectively.
Thus, each party $p \in \{A,B,C\}$ has qubits $\Q^p_1,\ldots,\Q^p_n$ 
and input string $x^p = x^p_1 \ldots x^p_n$, and the goal is for party 
$A$ to determine the value of $f(x^A,x^B,x^C)$.

The protocol begins by each party $p \in \{A,B,C\}$ performing the 
following operations and measurements on his qubits in order to 
obtain a bit $s^p$.\ee

\ni \ii {\bf for each $i \in \1n$ do} \nl
\ni \ii \ii {\bf if $x^p_i = 0$ then apply $H$ to} $\Q^p_i$ \nl
\ni \ii \ii {\bf measure} $\Q^p_i$ {\bf yielding bit $s^p_i$} \nl
\ni \ii $s^p \leftarrow s^p_1 + \cdots + s^p_n$\ee

\ni In the above, $H$ is the Hadamard transform, that maps 
$\ket{0}$ to ${1 \over \sqrt 2}(\ket{0}+\ket{1})$ and 
$\ket{1}$ to ${1 \over \sqrt 2}(\ket{0}-\ket{1})$ (and 
we recall that $+$ is in modulo two arithmetic).
Also, all measurements are in the standard basis consisting of 
$\ket{0}$ and $\ket{1}$.
Next, $B$ and $C$ send bits $s^B$ and $s^C$ respectively to $A$, 
who outputs the value of $s^A + s^B + s^C$.

This protocol works if and only if, for all $x^A, x^B, x^C \in \01^n$ 
such that $x^A + x^B + x^C = \one$, the bits $s^A$, $s^B$, $s^C$ satisfy 
\begin{equation}
s^A + s^B + s^C = f(x^A, x^B, x^C).
\label{property}
\end{equation}
The proof that Eq.~(\ref{property}) holds is based on the following lemma, 
which is equivalent to the result in \cite{Mermin}, though expressed in a 
different language.\ee

\loud{Lemma 1:}{\sl For all $i \in \1n$, 
\begin{equation}
s^A_i + s^B_i + s^C_i = x^A_i \o x^B_i \o x^C_i.
\end{equation}}\vspace*{-2mm}

\loud{Proof:} By Eq.~(\ref{conds}), 
$x^A_i x^B_i x^C_i \in \{001,010,100,111\}$.

First, consider the case where $x^A_i x^B_i x^C_i = 111$.
In this case, no $H$ transformation is applied to any of 
$\Q^A_i$, $\Q^B_i$, $\Q^C_i$.
Therefore, $\Q^A_i \Q^B_i \Q^C_i$ is measured in state (\ref{state}), 
which implies that $s^A_i + s^B_i + s^C_i = 1 = x^A_i \o x^B_i \o x^C_i$.

Next, in the case where $x^A_i x^B_i x^C_i = 001$, an $H$ transformation 
is applied to $\Q^A_i$ and to $\Q^B_i$ but not to $\Q^C_i$.
Therefore, $\Q^A_i \Q^B_i \Q^C_i$ is measured in state 
\begin{eqnarray}
\lefteqn{H \ox H \ox I \left( \half 
(\ket{001} + \ket{010} + \ket{100} - \ket{111})\right)} & & \nonumber \\
& = & \half (\ket{011} + \ket{101} + \ket{000} - \ket{110})
\end{eqnarray}
so $s^A_i + s^B_i + s^C_i = 0 = x^A_i \o x^B_i \o x^C_i$.
The cases where $x^A_i x^B_i x^C_i = 010$ and $100$ are similar by 
the symmetry of state (\ref{state}).$\Box$\ee

Now, it follows that 
\begin{eqnarray}
s^A + s^B + s^C & = & 
\left(\sum_{i=1}^n s^A_i\right) + \left(\sum_{i=1}^n s^B_i\right) 
+ \left(\sum_{i=1}^n s^C_i\right) \nonumber \\
& = & \sum_{i=1}^n (s^A_i + s^B_i + s^C_i) \nonumber \\
& = & \sum_{i=1}^n x^A_i \o x^B_i \o x^C_i \nonumber \\
& = & f(x^A, x^B, x^C).
\end{eqnarray}\ee

\loud{NO TWO-BIT CLASSICAL PROTOCOL \newline EXISTS:}\ee

We now show that, in the case where $n = 3$, without the 
use of entangled particles, two bits of communication among Alice, Bob, 
and Carol are insufficient for Alice to obtain enough information 
to deduce $f(x,y,z)$.
(This lower bound can be extended to all cases where $n > 3$ by fixing 
the value of all but the first $n$ inputs of each party.)

First, consider the possibilities of which parties the two bits are 
sent among.
Clearly there is no point in Alice sending the second bit.
Also, if Alice sends the first bit to, say, Bob then there is no point in 
Carol sending the second bit to Alice (since the first bit sent is then 
useless to Alice).
Therefore, if Alice sends the first bit to Bob then we can assume that 
Bob sends the second bit to Alice.
Also, note that, by substituting Eq.~(\ref{conds}) into Eq.~(\ref{fcn}), 
\begin{equation}
f(x,y,z) = x_1 \o y_1 +  x_2 \o y_2 + x_3 \o y_3.
\label{innerprod3}
\end{equation}
Thus, since only Alice and Bob are involved in the communication, 
this scenario reduces to the two-party inner product function, 
whose communication complexity is known to be three.
Therefore there is no protocol in which Alice sends one of the two bits 
to Bob.
Also, if Bob sends two bits to Alice then this can again be viewed 
as a two-bit two-party protocol computing (\ref{innerprod3}) which is 
impossible.
The above arguments also apply with Carol substituted for Bob.

The remaining possibilities are that Bob and Carol each send a single bit to 
Alice, or Bob sends a bit to Carol, who sends a bit to Alice (or vice versa).
Both of these are subsumed by the scenario where Bob is allowed to broadcast 
one bit to both Alice and Carol, and then Carol sends one bit to Alice, 
who must output $f(x,y,z)$.
This is the interesting case to examine.

The bit that Bob broadcasts is some function $\phi : \01^3 \rightarrow \01$ 
of his input data $y$ alone.
The function $\phi$ partitions $\01^3$ into two classes $\phi^{-1}(0)$ 
and $\phi^{-1}(1)$.
Call these two classes $S_0$ and $S_1$, and assume (without loss of 
generality) that $000 \in S_0$.
After Bob broadcasts his bit, what Alice and Carol each learn is whether 
$y \in S_0$ or $y \in S_1$.
For a two-bit protocol to be correct, it must always be possible at this 
stage for Carol to send one bit to Alice that will enable Alice to 
completely determine the value of $f(x,y,z)$.
We shall show that, whatever the partitioning $S_0, S_1$ is, there is an 
instance where Alice cannot determine the value of $f(x,y,z)$.
There are 128 different possible partitionings, and each is one of the 
seven types that are examined below.\ee

\loud{Case 1 {\boldmath $|S_0| \le 2$}:}
Recall our convention that $000 \in S_0$.
If $S_0$ has a second element then, by symmetry, no generality is lost if 
we assume that it is either $100$, $110$, or $111$.

Thus, without loss of generality, $001, 010, 011 \in S_1$.
Now, should the bit that Bob broadcasts specify to Alice and Carol 
that $y \in S_1$, Carol must send one bit to Alice from which Alice can 
completely determine the value of $f(x,y,z)$.
The bit that Carol sends induces a partition of the possible values 
of $z$ into two classes.
If $x = 001$ then, from Alice's perspective, after receiving Bob's 
bit but before receiving Carol's bit, the possible values of $(x,y,z)$ 
include $(001,001,111)$, $(001,010,100)$, $(001,011,101)$ 
and the respective values of $f(x,y,z)$ on these points are 1, 0, 1.
Therefore, for the protocol to be successful in this case, 
the partition that Carol's bit induces on $z$ must place $111$ and $101$ 
together in one class and $100$ in the other class (otherwise Alice would 
not be able to determine $f(x,y,z)$ when $x=001$).
On the other hand, if $x=011$ then, from Alice's perspective, the possible 
values of $(x,y,z)$ include 
$(011,001,101)$, $(011,010,110)$, $(011,011,111)$ and the respective 
values of $f(x,y,z)$ on these points are 1, 1, 0.
Since we have established that Carol's bit does not distinguish between 
$z = 111$ and $z = 101$, Carol's bit is not sufficient information for 
Alice to determine $f(x,y,z)$ in this case.\ee

\loud{Case 2 {\boldmath $|S_0| \ge 3$}:}
For this case, we consider the subcases where either $S_0$ contains 
a string of weight 1 (i.e. that has exactly one 1) or does not.\ee

\loud{Case 2.1 {\boldmath $|S_0|$ contains a string of weight 1}:}
Without loss of generality, assume $001 \in S_0$.
By our convention, $000 \in S_0$, and, after disregarding the obvious 
symmetries, there are four distinct possibilities for a third element of 
$S_0$: $010$, $011$, $110$, $111$ and these are considered separately.\ee

\loud{Case 2.1.1 {\boldmath $000, 001, 010 \in S_0$}:}
The argument is similar to that in Case 1 using $S_0$ instead of $S_1$.
Consider Alice's perspective.
If $x = 001$ then, the possible values for $(x,y,z)$ include 
$(001,000,110)$, $(001,001,111)$, $(001,010,100)$
for which the respective values of $f(x,y,z)$ are 0, 1, 0; 
whereas, if $x = 011$ then the 
possible values for $(x,y,z)$ include 
$(011,000,100)$, $(011,001,101)$, $(011,010,110)$
for which the respective values of $f(x,y,z)$ are 0, 1, 1.
No binary partitioning of $z$ will work for both possibilities.\ee

\loud{Case 2.1.2 {\boldmath $000, 001, 011 \in S_0$}:}
Consider Alice's perspective.
If $x = 001$ then, the possible values for $(x,y,z)$ include 
$(001,000,110)$, $(001,001,111)$, $(001,011,101)$ for which the respective 
values of $f(x,y,z)$ are 0, 1, 1; whereas, if $x = 011$ then the 
possible values for $(x,y,z)$ include 
$(011,000,100)$, $(011,001,101)$, $(011,011,111)$ 
for which the respective values of $f(x,y,z)$ are 0, 1, 0.
No binary partitioning of $z$ will work for both possibilities.\ee

\loud{Case 2.1.3 {\boldmath $000, 001, 110 \in S_0$}:}
Consider Alice's perspective.
If $x = 010$ then, the possible values for $(x,y,z)$ include 
$(010,000,101)$, $(010,001,100)$, $(010,110,011)$ for which the respective 
values of $f(x,y,z)$ are 0, 0, 1; whereas, if $x = 011$ then the 
possible values for $(x,y,z)$ include 
$(011,000,100)$, $(011,001,101)$, $(011,110,010)$ 
for which the respective values of $f(x,y,z)$ are 0, 1, 1.
No binary partitioning of $z$ will work for both possibilities.\ee

\loud{Case 2.1.4 {\boldmath $000, 001, 111 \in S_0$}:}
Consider Alice's perspective.
If $x = 010$ then, the possible values for $(x,y,z)$ include 
$(010,000,101)$, $(010,001,100)$, $(010,111,010)$ for which the respective 
values of $f(x,y,z)$ are 0, 0, 1; whereas, if $x = 011$ then the 
possible values for $(x,y,z)$ include 
$(011,000,100)$, $(011,001,101)$, $(011,111,011)$ 
for which the respective values of $f(x,y,z)$ are 0, 1, 0.
No binary partitioning of $z$ will work for both possibilities.\ee

\loud{Case 2.2 {\boldmath $|S_0|$ contains no string of weight 1}:}
We consider the following three subcases.\ee

\loud{Case 2.2.1 {\boldmath $111 \not\in S_0$}:}
In this case, $001, 010, 100, 111 \in S_1$.
Suppose that Bob's bit specifies that $y \in S_1$.
Consider Alice's perspective.
If $x = 001$ then, the possible values for $(x,y,z)$ include 
$(001,001,111)$, $(001,010,100)$, $(001,100,010)$, $(001,111,001)$ 
for which the respective 
values of $f(x,y,z)$ are 1, 0, 0, 1; whereas, if $x = 010$ then the 
possible values for $(x,y,z)$ include 
$(010,001,100)$, $(010,010,111)$, $(010,100,001)$, $(010,111,010)$ 
for which the respective values of $f(x,y,z)$ are 0, 1, 0, 1.
No binary partitioning of $z$ will work for both possibilities.\ee

\loud{Case 2.2.2 {\boldmath $111 \in S_0$}:}
In this case, $S_0$ must contain an element of weight 2.
Without loss of generality, $011 \in S_0$.
Therefore, $000, 011, 111 \in S_0$.
Consider Alice's perspective.
If $x = 010$ then, the possible values for $(x,y,z)$ include 
$(010,000,101)$, $(010,011,110)$, $(010,111,010)$ for which the respective 
values of $f(x,y,z)$ are 0, 1, 1; whereas, if $x = 110$ then the 
possible values for $(x,y,z)$ include 
$(110,000,001)$, $(110,011,010)$, $(110,111,110)$ 
for which the respective values of $f(x,y,z)$ are 0, 1, 0.
No binary partitioning of $z$ will work for both possibilities.\ee

This concludes the proof that there is no classical protocol for computing 
$f(x,y,z)$ in which only two bits are communicated among Alice, Bob, 
and Carol.\ee

\loud{A THREE-BIT CLASSICAL PROTOCOL:}\ee

Although one might suspect that, without the use of entangled particles, 
$n$ bits of communication are necessary for Alice to determine $f(x,y,z)$ 
in general, it turns out that three bits always suffice.

The idea behind the method is to count the total number of 0s among all 
the $3n$ inputs of Alice, Bob, and Carol.
Note that, for each $i \in \{1,\ldots,n\}$, if $x_i \o y_i \o y_i = 1$ then 
there are zero 0s among $x_i, y_i, z_i$, and if $x_i \o y_i \o y_i = 0$ 
then there are two 0s among $x_i, y_i, z_i$.
Let the number of 0s among $x_1,\ldots,x_n$ be $r_A$, the number of 0s 
among $y_1,\ldots,y_n$ be $r_B$, and the number of 0s among 
$z_1,\ldots,z_n$ be $r_C$.
Let $k$ be the total number of terms among 
$x_1 \o y_1 \o z_1, \ldots,  x_n \o y_n \o z_n$ that have value 0.
Then, from the above, 
$r_A + r_B + r_C = 2 k$.
Therefore, it suffices for Bob to send $r_B$ to Alice and Carol to 
send $r_C$ to Alice in order for Alice to compute $k$.
From $k$, Alice can easily compute $f(x,y,z) = (n - k) \bmod 2$.
This involves $2 \log n$ bits of communication.
Fortnow \cite{Fortnow} has shown that the communication can be 
reduced to three bits as follows.
Since Alice only needs the parity of $k$, she only needs 
the values of $r_A$, $r_B$, $r_C$ in modulo 4 arithmetic.
Therefore, it suffices for Bob and Carol to each send two bits to Alice.
This yields a four-bit protocol.
To obtain a three-bit protocol, note that $r_A + r_B + r_C$ 
is guaranteed to be an even number.
This means that either Bob or Carol can send just 
the high order bit of his/her two-bit number.\ee

We would like to thank Lance Fortnow for allowing us to incorporate 
his improvement to the $2\log n$-bit classical protocol.
R.C. is grateful to Lev Vaidman for introducing him to the three-particle 
deterministic examples of Bell nonlocality, Richard Jozsa for 
providing references to this work, and David DiVincenzo for general 
discussions about the power of entanglement.
R.C. is supported in part by Canada's NSERC.
H.B. is supported by: NWO by SION Project 612-34-002, EU through 
NeuroCOLT ESPRIT Working Group Nr.\ 8556, and HC\&M grant CCR 92-09833.

\end{multicols}

\end{document}